\documentstyle{article}
\newcommand{\tr}{\mathop{\rm tr}\nolimits}
\newcommand{\Tr}{\mathop{\rm Tr}\nolimits}
\begin{document}
\title{Anomalies in nonrelativistic quantum mechanics}
\author{D.A.Kirzhnits and G.V.Shpatakovskaya\\
Institute for Mathematical Modelling, RAS}
\maketitle
\begin{abstract}
It is shown that if a potential in a  nonrelativistic system
of Fermi particles has a sufficiently strong singularity, anomalies
(nonzero values of quantities formally equal to zero) will probably
appear. For different types of singularities (in paticular, for the
Coulomb potential), anomalies associated with the energy and total
number of particles in the system are calculated. These anomalies
may be beneficial in deriving a semiclassical description of electron-
nuclear systems.
\end{abstract}
        As is known (see, e.g.,\cite{Kafiev-91}), the subject of anomalies
is discussed in situations where quantities equal to zero in the classical
limit, thanks to their being invariant, acquire some finite value due to
quantum effects in the high-energy region (and, in general, within their
singularity domain) The known examples of anomalies (including the
historically first example of the appearance of a diamagnetic, proportional
to the potential and, as a result, the gauge invariance-violating component
of the current in the theory of vacuum polarization) are from quantum field
theory. Anomalies in nonrelativistic quantum mechanics are discussed below.
When used, these anomalies simplify the description of systems of interacting
particles \cite{Kirzhnits-94}.

        {\bf 1.} The number $N$ and the energy $E$ of Fermi particles at a
nonzero temperature $T$ in an external field $U$ are defined by \footnote
{The results obtained below also appear to be true in the case of accounting
for interactions between particles by the method of self-consistent fields
(Hartee, Hartree-Fock).}
$$
            N = 2\tr(\hat{\rho}),\quad
            E = 2\tr(\hat{H}\hat{\rho})
$$
where $\hat{H}=\hat{p}^2/2m + U$ is the Hamiltonian, $\hat{\rho}=\theta(\mu-
\hat{H})$ is the level-filling operator with an upper bound $\mu$,
($\theta(x)=1$ for $x>0$ and $\theta(x)=0$ for $x<0$),

$$\tr(...) = (2\pi\hbar)^{-3}\int d\vec{x}d\vec{p}
\exp(-i\vec{x}\vec{p}/\hbar)(...)\exp(i\vec{x}\vec{p}/\hbar)$$

        If $\mu>0$, in oder to do without both continuous and discrete
spectra regims, as well as for other reasons (see \cite{Kirzhnits-94},
\cite{Baz-71}), it is convenient to pass to
\begin{equation}
     N = a_N - 2\tr(\hat{\tilde{\rho}}),\quad
        E =a_E - 2\tr(\hat{H}\hat{\tilde{\rho}}),
\end{equation}
where $\hat{\tilde{\rho}}=1-\hat{\rho}$ is the level-filling operator
with a low limit $\mu$,
\begin{equation}
     a_N = 2\tr(1),\quad
        a_E = 2\tr(\hat{H}).
\end{equation}
        As can be immediately seen, the dependence of $a_N$ and $a_E$ on the
Planck constant $\hbar$ is concentrated in $(2\pi\hbar)^{-3}$. As a result,
quantities $A_{\alpha}=(2\pi\hbar)^3 a_{\alpha},\quad \alpha=N,E $, do not
depend on $\hbar$, i.e. are invariant under the transformation
        $$\hbar\to\hbar+C.$$
Introducing $A^0_\alpha = \lim_{\hbar\to 0}A_\alpha,\quad \alpha=N,E $
(which fit the Thomas-Fermi model where the commutators of  operators
$\hat{p}^2/2m$ and $U$ are equal to zero), we conclude that
\begin{equation}
     \delta A_\alpha = A_\alpha - A^0_\alpha=0,\quad \alpha=N,E
\end{equation}
These differences represent the total sum of quantum and shell corrections to
the  Thomas-Fermi model \cite{Kirzhnits-94} for the corresponding values of
$A_\alpha$.

        {\bf 2.} Equalities (3) have a formal meaning and can be violated due
to strong divergences of $A_\alpha$ and $A^0_\alpha$ at large $p$. For example,
this occurs in the case of a sufficiently strong sigularity of $U$ at small $x$.
The corresponding $\delta A_\alpha$ are examples of quantum-mechanical anomalies.

        To suppress the contribution from the region of large $p$ at an
intermediate stage, let us introduce the regularization
$$
\delta A_N\to\lim_{\Lambda\to\infty}2\Lambda^2
  \left[\{\Tr(\Lambda+\hat{H})^{-2}\}-\{\hbar=0\}\right]=
-\lim_{\Lambda\to\infty} 2\Lambda^2\frac{\partial W}{\partial\Lambda},\eqno (4a)
$$
$$
\delta A_E \to \lim_{\Lambda\to\infty}2\Lambda^2
  \left[\{\Tr(\hat{H}(\Lambda+\hat{H})^{-2})\}-\{\hbar=0\}\right]=
\lim_{\Lambda\to\infty}2\Lambda^2
\left(1+\Lambda\frac{\partial}{\partial\Lambda}\right)W,\eqno (4b)
\setcounter{equation}{4}
$$
where  $\Tr=(2\pi\hbar)^3\tr $,
\begin{eqnarray}
  W=\left[
\{\Tr(\Lambda+\hat{H})^{-1}\}-\{\hbar=0\}
\right]=\nonumber\\
\int d\vec{x}d\vec{p}
\left[
(\Lambda+(\vec{p}-i\hbar\nabla)^2/2m+U)^{-1}-
  (\Lambda+p^2/2m+U)^{-1}
\right]
\end{eqnarray}
In Eqs. (4) and (5), the second braces contain the same quantities as the first,
though at $\hbar=0$, and the gradient in Eq.(5) acts on U.

        Replacing $\vec{p}\to\Lambda^{1/2}\vec{p},\quad\vec{x}\to\Lambda^{-1/2}
\vec{x},\mbox{ and }U(\vec{x})\to U(\Lambda^{-1/2}\vec{x})$, we can see that the
relative contribution of the potential $U$ to Eq.(5) is determined by the
behavior of the potential as $x\to 0$ (at large energies). If $U$ has a
singularity of type $x^{-2}$ at this point, then $W\propto\Lambda^{-1}$ and
all perturbative orders in $U$ are equally important. At the same time,
$\delta A_{N}$  has a finite value and, according to Eq.(4b), $\delta A_E$ equals
 zero (case A). If, at small $x$, the potential like a Coulomb one,
corrections of the first ($W\propto\Lambda^{-3/2}$) and second
($W\propto\Lambda^{-2}$) orders in the perturbation theory are importent and,
at the same time, $\delta A_N = 0$ and $\delta A_E$ differs from zero (case B).
Finally, if the singularities of $U$ at small $x$ are weaker than the Coulomb
singularity, the first order of perturbation theory, at most, is important
(case C).

      {\bf 3.} From this point on, we use the well-known expansion to calculate (5),
$$
(a+b)^{-1}=a^{-1}-a^{-1}ba^{-1}+a^{-1}ba^{-1}ba^{-1}+... .
$$
To begin with, we consider the first order of perturbation theory that can be
related to all three of the above-mentioned cases:

\begin{eqnarray*}
W_1=-\int d\vec{p} \int d\vec{k}U(\vec{k})\delta(\vec{k})
\bigl[
(\Lambda+(\vec{p}+\vec{k})^2/2m )^{-1}-\\
(\Lambda+p^2/2m )^{-1}
\bigr]
(\Lambda+p^2/2m )^{-1}
\end{eqnarray*}

After being averaged over angles, the expression in square brackets behaves
as $k^2$ as $ k\to 0 $. As a result, $W_1$ is different from zero only if
$U(k)$ has a sufficiently strong singularity, e.g., of type $\propto k_{-2}$
as $k \to 0$. This means that the potential $U(x)$ has the Coulomb behavior
$Ze^2/x$ at large $x$. Taking advantage of the Feynman formula

\begin{equation}
(ab)^{-1}=\int\limits_{0}^{1}dx [ax+b(1-x)]^{-2},
\end{equation}

   it is easy to find that
$$
W_1 \propto (2\pi\hbar)^3 Z e^2 m^2 /[\hbar(2m\Lambda)^{3/2}].
$$
Accordingly, $\delta A_N=0 $ and $\delta A_E $ is different from zero and is
infinite,
$$
\delta A_E \propto (2\pi\hbar)^3 Z e^2 (2m\Lambda)^{1/2}/\hbar.
$$

This result pertains to the case of Coulomb behavior at large distances (an
unscreened system) and is of little interest from the physical point of view.
In what follows, we remove this case from consideration, as well as the case
of a weaker-than-Coulomb decrease at large $x$ and, thus, avoid the necessity
of discussing the first order of perturbation theory.

{\bf 4.} Let us pass to the discussion of case A ($U\to \alpha x^{-2}$ at
small $x$ and restrict ourselves to the calculation of $\delta A_N $ different
from zero. Replacing $\vec p\to\sqrt{2m\Lambda}\vec p,\quad
\vec x\to\sqrt{\alpha/\Lambda}\vec x,$ from Eqs. (4) and (5), we obtain
\begin{eqnarray*}
\delta A_N=-2(2m\alpha)^{3/2}\int d\vec{x}d\vec{p}\bigl[(1+(\vec{p}-
i\hbar\nabla/\sqrt{2m\Lambda})^2+1/x^2)^{-1}- \\
(1+p^2+1/x^2)^{-1}\bigr].
\end{eqnarray*}
We assume that the interaction is strong ($m\alpha/\hbar^2>>1$) and restrict
ourselves to the first unvanishing term in the expansion in terms of
$\alpha^{-1}$,
\begin{equation}
\delta A_N=-\frac{(2\pi\hbar)^3}{36}\sqrt{2m\alpha}/\hbar
\end{equation}
It should be noted that such an anomaly reflects, in a definite sense, the
violation of completeness of the system of eigenfunctions of the
Schr\"odinger equation with singular potential $U$. Indeed, accounting for
the completeness condition
$$
\sum_{\nu}\bar{\psi}_{\nu}(\vec{x})\psi_{\nu}(\vec{x})=\delta(\vec{x}-\vec{x}')
$$
and writing $a_N$ in the form
$$
a_N=2\tr(1)=2\sum_{\nu}\int d\vec{x}\bar{\psi}_{\nu}(\vec{x})
\psi_{\nu}(\vec{x})=2\int d\vec{x}\delta(0)
$$
we come to the formal conclusion of the independence of $a_N=A_N/(2\pi\hbar)^3$
from $\hbar$. The presence of divergences at large energies strips this formal
conclusion of meaning (see Eq.(7)).

  {\bf 5.} In this concluding section, the quantity $\delta A_E$ is discussed,
being of great physical interest in the case for the Coulomb behavior
$U\propto -Z/r $ at small distances (case B). Restricting the discussion
to the second order of perturbation theory in accordance with what has
been said above, we have
\begin{eqnarray}
W_2 = (2\pi\hbar)^{-3}\int d\vec{p}\int d\vec{k}|U(\vec{k})|^2
\big[
(\Lambda+
(\vec{p}+\vec{k})^2/2m)^{-1}-\nonumber\\
(\Lambda+p^2/2m)^{-1}
\bigr]
(\Lambda+p^2/2m)^{-2}
\end{eqnarray}
where $U(\vec{k})=4\pi Ze^2\hbar^2/k^2$. Presenting the last factor in Eq.(8)
in the form
$$
-\frac{\partial}{\partial\Lambda'}
      \left.\left(\Lambda'+\frac{p^2}{2m}\right)^{-1}\right|_{\Lambda'=\Lambda}
$$
and taking advantage of Eq.(6), we obtain
$$
W_2=-(2\pi\hbar)^3\frac{Z^2}{8\Lambda^2}\cdot\frac{e^2}{a_0}.
$$
According to Eq.(3), this leads to the finite value
$$
\delta A_E=(2\pi\hbar)^3\frac{Z^2}{4}\cdot\frac{e^2}{a_0}
$$
previously given without proof by the authors in [2].

{99}
\end{document}